\documentclass[12pt]{iopart}

\usepackage{graphicx}
\begin{document}
\title{Coexisting tuneable fractions of glassy and equilibrium long-range-order phases in manganites.}
\author{A. Banerjee, A. K. Pramanik, Kranti Kumar and P. Chaddah.}
\address{UGC-DAE Consortium for Scientific Research,\\ University Campus, Khandwa Road, Indore-452017, INDIA.}
\ead{alok@csr.ernet.in}
\begin{abstract}
Antiferromagnetic-insulating(AF-I) and the ferromagnetic-metallic(FM-M) phases coexist in various half-doped manganites over a range of temperature and magnetic field, and this is often believed to be an essential ingredient to their colossal magnetoresistence. We present magnetization and resistivity measurements on Pr$_{0.5}$Ca$_{0.5}$Mn$_{0.975}$Al$_{0.025}$O$_3$ and Pr$_{0.5}$Sr$_{0.5}$MnO$_{3}$ showing that the fraction of the two coexisting phases at low-temperature in any specified measuring field H, can be continuously controlled by following designed protocols traversing field-temperature space; for both materials the FM-M fraction rises under similar cooling paths. Constant-field temperature variations however show that the former sample undergoes a 1st order transition from AF-I to FM-M with decreasing T, while the latter undergoes the reverse transition. We suggest that the observed path-dependent phase-separated states result from the low-T equilibrium phase coexisting with supercooled glass-like high temperature phase, where the low-T equilibrium phases are actually homogeneous FM-M and AF-I phases respectively  for the two materials.
\end{abstract}

\maketitle
Half-doped perovskite manganites with general formula (R$_{0.5}$A$_{0.5}$)MnO$_3$ (where R is rare-earth and A is alkaline-earth elements) belong to a class of materials where transformation from antiferromagnetic (AF)-insulating (I) to ferromagnetic (FM)- metallic (M) phase can be triggered by variation of temperature and magnetic field \cite{Dag,Tokura1,Kuwa}. The symmetrically incompatible competing orders are separated by a first order phase transition (FOPT) that may be caused by varying control variables like temperature or field, triggering giant response in some physical properties, magnetoresistance being the most prominent of these. The half doped manganites are intrinsically disordered because of randomness in R/A site. Early theoretical work predicted that such quenched disorder would give rise to a landscape of free-energies, resulting in spatial distribution of the coexisting phases \cite{Imry}; strain-induced phase coexistence has also been proposed in manganites \cite{Ahn}. Such disorders endow a finite broadening to the transition causing a coexistence of two phases with different order parameters \cite{Chat}. Existence of such a spatial distribution of coexisting phases, and its alteration with the control variables across disorder-broadened FOPT in diverse systems have been demonstrated using a variety of probes \cite{Cheong, Soibel, Mathur, Roy, DD, Park, Wu}. Outside the broadened transition region, a canonical system recovers the respective homogeneous phases completely, and attains equilibrium. However, this inhomogeneous phase coexistence persists to the lowest temperature in many materials \cite{Kuwa, Pecharsky, Chat2, Roy2}, leading to theoretical work proposing interesting equilibrium phases \cite{Littlewood}. Often the initial state is not recovered after going through a full cycle of the control variable \cite{Chat2,Roy2,Manekar}. It has been shown in many manganites around half doping that once AF-I phase is converted to FM-M phase by the application of external magnetic field at low temperatures, the system does not go back to starting phase when the field is withdrawn \cite{Tokura1, Kuwa}. Similar asymmetry is observed recently with the cycling of temperature in fixed field \cite{Banerjee, Rawat}. These give clear evidence that the reverse transformation is hindered and the initial zero-field cooled (ZFC) state is not recovered completely, resulting in a zero field state with two coexisting phases. 

           The so-called 'robust' charge ordered manganites with narrow one-electron bandwidth like Pr$_{0.5}$Ca$_{0.5}$MnO$_{3}$ or Nd$_{0.5}$Ca$_{0.5}$MnO$_{3}$ show AF-I to FM-M transition in high pulsed magnetic field ($>$200 kOe) \cite{Toku}. However, when the charge ordering is weakened by small substitution in the Mn-site of Pr$_{0.5}$Ca$_{0.5}$MnO$_{3}$, the field induced AF-I to FM-M transition occurs at accessible static field ranges and show striking effects \cite{Mahe, Sunil, Banerjee}. We present here high-field magnetization and resistivity studies on Pr$_{0.5}$Ca$_{0.5}$Mn$_{0.975}$Al$_{0.025}$O$_3$ and the wider bandwidth compound Pr$_{0.5}$Sr$_{0.5}$MnO$_3$. We show that these functional properties, at a specified field at low-T, can be tuned by following novel paths in H-T space. We argue that the observed path-dependent phase-separated states are glass-like states where kinetics is arrested. Conventional measurements mask the simple equilibrium phase diagram -- the arrest of dynamics masks thermodynamics. These results have general applicability to first-order magnetic transition in systems with intrinsic disorder, including intermetallic alloys, which show glass-like freezing of higher-temperature long-range-order phase \cite{Chat,Manekar}. Moreover, similar studies on such materials would shed light on the physics of glass formation since magnetic field can be tuned easily compared to external pressure used for conventional studies \cite{Stanley}.
  
        Polycrystalline sample of Pr$_{0.5}$Ca$_{0.5}$Mn$_{0.975}$Al$_{0.025}$O$_3$ was prepared by solid-state reaction. Details of the sample preparation and characterization can be found in Ref. [25]. Polycrystalline sample of Pr$_{0.5}$Sr$_{0.5}$MnO$_3$ was prepared by solid-state route with starting materials having purity at least 99.99\%. The x-ray diffraction (XRD) measurements are done using a Rigaku Rotaflex RTC 300 RC powder diffractometer with Cu K$\alpha$ radiation from rotating anode x-ray generator (18 kW). The XRD data is analyzed by Rietveld profile refinement and samples are found to be in single crystallographic phase, without any detectable impurity phase. Iodometric redox titration is performed to estimate the Mn$^{3+}$/Mn$^{4+}$ ratio. The samples are found to be half doped within the detection limit. The resistivity and magnetic measurements are performed using commercial set-ups (14Tesla-PPMS-VSM, M/s. Quantum Design, USA).

      Figures 1a and 1b show data at 5K for Pr$_{0.5}$Ca$_{0.5}$Mn$_{0.975}$Al$_{0.025}$O$_3$. The isothermal M-H at 5 K shows that ZFC AF state transforms to FM state with increase in field above 55 kOe, but the initial AF state does not recover when the field is reduced from 140 kOe to zero. Subsequent H-increasing curve shows no signature of the AF-to-FM transition, indicating that the remanent state is fully FM. Similarly the resistivity vs H measured at 5 K after ZFC shows that the resistance, which was above our measurement limit at H=0, becomes measurable and decreases by orders of magnitude with the increase in H indicating an insulator to metal like change. However, the initial high resistance value is not recovered when the field is reduced to zero, implying the stabilization of the metallic phase once it is produced. This is similar to the magnetization data, and there is no further conversion with field cycling. These two measurements at 5 K show that the H-induced FM-M state is (meta)stable under large excursions in H at 5 K. Softening of this anomaly is observed at 40 K in fig. 1(c) indicating part conversion to the ZFC AF-I state. Qualitatively similar behavior is observed at 5 K for other half doped manganites having wider bandwidth than Pr$_{0.5}$Ca$_{0.5}$MnO$_3$ and different spin, orbital and charge order like Pr$_{0.5}$Sr$_{0.5}$MnO$_3$ as shown in fig. 1d. For all these cases, the virgin M-H or R-H curves are not traced in subsequent field increasing cycles. The original homogeneous ZFC state transforms into an inhomogeneous state and shows phase coexistence even in zero field. It is remarkable that such anomalies are also observed in variety of systems like doped intermetallic alloys or giant magnetocaloric compound and is attributed to the interruption of the first-order transformation process by critically slow dynamics similar to glass like freezing of the supercooled high temperature phase \cite{Chat,Pecharsky,Chat2,Roy2,Manekar}. This phenomenon of hindered transformation or 'kinetic arrest' is also reported in 'charge ordered' La-Pr-Ca-Mn-O \cite{Kumar}. 

       More vivid manifestation of this non-ergodicity are portrayed in fig. 2 which show different path-dependent magnetization and conductivity values at same temperature and field for both Pr$_{0.5}$Ca$_{0.5}$Mn$_{0.975}$Al$_{0.025}$O$_3$ and Pr$_{0.5}$Sr$_{0.5}$MnO$_3$. Magnetization and resistivity data is obtained at 5 K after cooling each time from 320 K to 5 K, but in different cooling fields. The field is then varied isothermally to the specified measuring value. Data is shown in fig. 2 for two values of measuring field, and various values of cooling field. Data has been collected for many more values of measuring field, and shows similar behaviour. We find the general result for both samples that cooling in high field gives an increasing FM-M component, and this appears to be a generic feature of manganites around half doping. These path-dependent values clearly show the tuneable coexistence of distinct electronic and magnetic phases.

     We have shown in fig 2 that observed physical properties depend drastically but controllably on the field-temperature path followed to reach the measuring field and temperature. This presents a novel ability to control functional properties. We now show that in spite of this apparent commonality between these two different half doped systems the underlying FOPT is very different. Figures 3a and 3b show the thermal hysteresis in magnetization, indicating a broad hysteretic first-order transition, in 10 kOe field for Pr$_{0.5}$Ca$_{0.5}$Mn$_{0.975}$Al$_{0.025}$O$_3$ and Pr$_{0.5}$Sr$_{0.5}$MnO$_3$ respectively. We mention here that the conductivity also shows signatures of the corresponding FOPT. Thus the complex low temperature behaviour of these two half-doped manganites in field-temperature plane can be categorized in two groups. For one group [fig. 3(a)], the high-temperature phase is AF-I whereas for other [fig. 3(b)] it is FM-M. In spite of these contrasting 1st order transitions, the path-dependent effects depicted in figures 1 and 2 are similar. 

    To develop some unified understanding about the complex behaviour of field-temperature induced coexisting phases from simple representations, we redraw the recently proposed heuristic phase diagram for the two groups \cite {Banerjee, Rawat, Kumar}. The disorder broadening of FOPT results in broadening of the supercooling (H*, T*) and superheating (H**, T**) spinodals to bands. The kinetic arrest akin to the glass transition line would also broaden into (H$_K$, T$_K$) band because of disorder \cite{Manekar}. The sign of the slopes of (H*, T*) and (H**, T**) bands are decided from the fact that the ferromagnetic phase exists over larger temperature range in higher fields. The same for the (H$_K$, T$_K$) band is decided by the fact that if increase in temperature causes de-arrest to AF (FM) state, then isothermal de-arrest must be observed on lowering (raising) the field. The diagram of Fig. 3c is for the case showing first-order transformation from higher temperature AF phase to FM at low temperature. Materials like Pr$_{0.5}$Ca$_{0.5}$Mn$_{0.975}$Al$_{0.025}$O$_3$, following fig. 3c, would show a homogeneous AF phase on cooling in zero field, completely masking the FOPT to the FM phase. This is the case for the magnetocaloric Gd$_5$Ge$_4$ \cite{Roy2}. We then obtain from the phase diagram that cooling in fields above H$_2$ results in full transformation to equilibrium FM-M phase, whereas cooling in field below H$_1$ results in completely frozen AF-I state. Cooling in fields between H$_2$ and H$_1$ can generate a continuum of states with varying faction of metastable AF-I phase. The AF-I fraction remains constant in subsequent isothermal field variations provided one stays below the (H$_K$, T$_K$) band, explaining the data in figures 2a and 2b. Measurements following unusual paths in the temperature-field plane are essential to unravel the zero field equilibrium FM phase. Figure 3d shows the H-T schematic for the case when high temperature FM phase transforms to AF phase at low temperature and represents systems like Pr$_{0.5}$Sr$_{0.5}$MnO$_3$, Nd$_{0.5}$Sr$_{0.5}$MnO$_3$ \cite{Rawat} etc. Thus Figs. 3c and 3d encompasses wide range of half doped manganites and also other systems which show similar field temperature induced first-order transition. 

     The salient features of the proposed phase diagrams thus are: (i) Cooling across (H$_K$, T$_K$) band would freeze the supercooled higher temperature phase; (ii) Warming across this would de-arrest the frozen phase and (iii)Traversing this band in similar sense at a fixed temperature would accordingly result in arrest or de-arrest of the higher temperature phase. The significant outcome of Figs. 3c and 3d are that cooling the system between fields H$_1$ and H$_2$ can produce a continuum of coexisting phase fractions. Cooling in fields below H$_1$ or above H$_2$ results in two different homogeneous phases. One of which is in equilibrium and the other is in frozen metastable state, but both persist down to the lowest temperature. For the cases of both figs. 3c and 3d the coexisting phase fractions can be tuned in a qualitatively similar manner, but the behavior of these states on warming are contrasting. Figs. 3e and 3f show the magnetization behaviour for Pr$_{0.5}$Ca$_{0.5}$Mn$_{0.975}$Al$_{0.025}$O$_3$ and Pr$_{0.5}$Sr$_{0.5}$MnO$_3$ respectively, while warming in a fixed field (H$_W$) after cooling to 5 K in different fields (H$_C$). Larger fraction of metastable higher-temperature phase is frozen in Pr$_{0.5}$Ca$_{0.5}$Mn$_{0.975}$Al$_{0.025}$O$_3$ (Pr$_{0.5}$Sr$_{0.5}$MnO$_3$) while cooling in lower (higher) fields. We observe two sharp changes of opposite sign during warming when [H$_W$ - H$_C$] is positive (negative), and this is ascribed to de-arrest of the frozen phase at the (H$_K$, T$_K$) band, followed by the FOPT of the equilibrium low-temperature phase at the superheating band. On the other hand, cooling in higher (lower) fields for Pr$_{0.5}$Ca$_{0.5}$Mn$_{0.975}$Al$_{0.025}$O$_3$ (Pr$_{0.5}$Sr$_{0.5}$MnO$_3$) gives lower fraction of metastable frozen phase, and the sharp change signifying de-arrest does not become observable in the measurement time scale. This is a detail, which correlates the broadening of the (H*, T*) and the (H$_K$, T$_K$) bands \cite{Kumar}. It should be noted that the lack of de-arrest reported here is as observed on short time scale, and does not imply that the system is in equilibrium. Long-time relaxation is seen as expected for all the metastable states \cite{Chat2,Roy2,Banerjee}. 

    To conclude, our measurements show that the lack of dynamics triumphs over thermodynamics at the magnetic FOPT in the sample studied, resulting in an apparently complex phase diagram. The tangled web \cite{Kuwa,Littlewood} can be unraveled by extensively traversing various paths in H-T plane as per our protocol \cite{Banerjee, Kumar}, to obtain the equilibrium phase diagram; zero-field measurements or isothermal field variations (as in fig. 1) are not conclusive. We point out that intrinsic disorders including strains \cite{Ahn} can broaden the supercooling band and cause a finite overlap with the (H$_K$, T$_K$) band at zero field. This would give rise to coexisting phases even on ZFC for both figs. 3c and 3d. These phase diagrams can qualitatively explain various anomalies observed not just in half-doped manganites but also in a variety of magnetocaloric, martensitic, and other intermetallic materials. We hope that our conclusions would be vindicated by more direct mesoscopic measurements like micro-Hall probe scans, magnetic-force microscopy, XRD, etc which would have to be performed in high magnetic field.

		We benefited from discussion with S. B. Roy. DST, Government of India is acknowledged for funding the 14 Tesla-PPMS-VSM. A.K.P. acknowledges CSIR, India for fellowship.

\newpage

\begin{figure}
	\centering
		\includegraphics{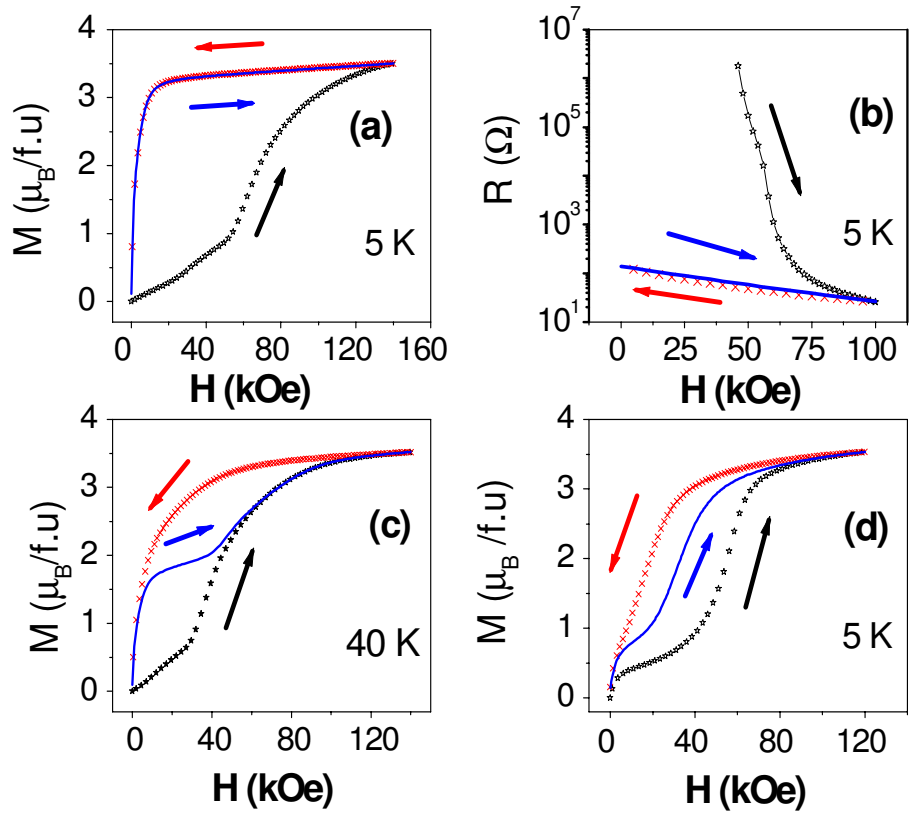}
	\caption{Isothermal magnetic field (H) dependence of magnetization (M) and resistivity (R) of the zero-field cooled (ZFC) state. (a) M vs. H of Pr$_{0.5}$Ca$_{0.5}$Mn$_{0.975}$Al$_{0.025}$O$_3$ at 5 K. After reaching the highest field, the magnetization of the field reducing path shows a complete overlap with the subsequent field increasing path. (b) The resistivity of Pr$_{0.5}$Ca$_{0.5}$Mn$_{0.975}$Al$_{0.025}$O$_3$ after ZFC at 5 K becomes measurable above 50 kOe. The virgin field increasing path is not traced in the following field cycles, similar to the magnetization (panel a) where there is no further conversion with field cycling. (c) The initial or virgin M vs. H curve of Pr$_{0.5}$Ca$_{0.5}$Mn$_{0.975}$Al$_{0.025}$O$_3$ at 40 K shows the decrease in the field (~30 kOe) of the AF to FM conversion. The subsequent field decreasing and increasing paths (which define the envelope curve) do not overlap (unlike at 5 K) but lie above the virgin curve indicating the coexisting AF and FM phases. (d) The M vs. H curve of Pr$_{0.5}$Sr$_{0.5}$MnO$_3$ at 5 K has feature similar to panel (c) where the virgin curve lie below the envelope curve during subsequent field cycling giving rise to irrecoverable field induced transition.}
	\label{fig:Fig1}
\end{figure}

\begin{figure}
	\centering
		\includegraphics{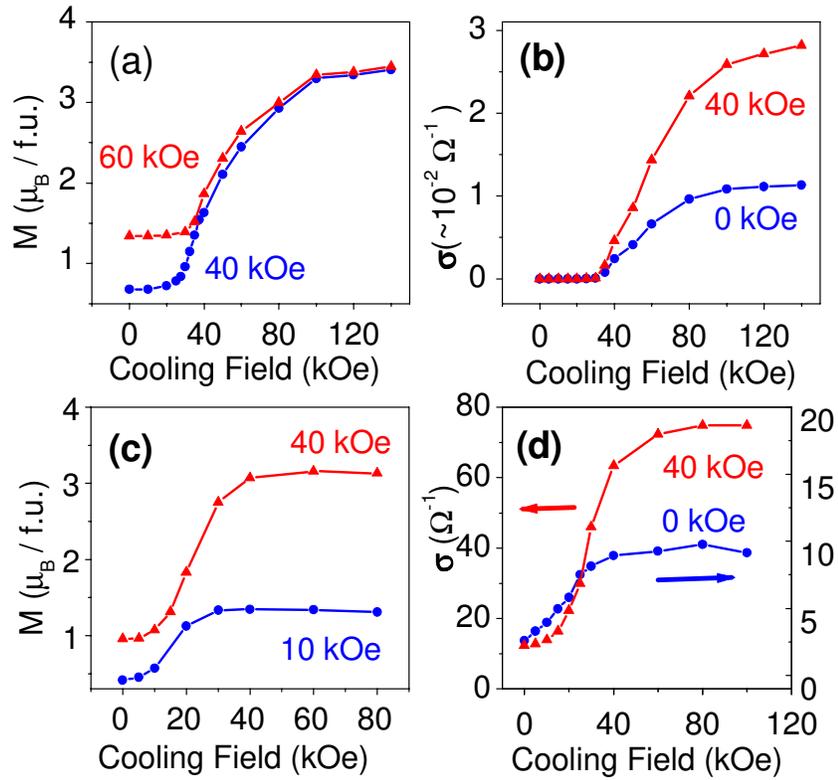}
	\caption{a, Pr$_{0.5}$Ca$_{0.5}$Mn$_{0.975}$Al$_{0.025}$O$_3$  is cooled each time from 320 K to 5 K in different cooling fields, and then the magnetization is measured after reaching specified measuring field isothermally at 5 K. Data is shown for measuring fields being 60 kOe (red triangles) and 40 kOe (blue circles). b, Conductivity for Pr$_{0.5}$Ca$_{0.5}$Mn$_{0.975}$Al$_{0.025}$O$_3$ taken under similar protocols, with measuring field being zero (blue circles) and 40 kOe (red triangle). c, Different values of magnetization at fixed temperature and field after cooling in different fields for Pr$_{0.5}$Sr$_{0.5}$MnO$_3$  similar to the panel a. In this case the measurement fields are 10 kOe (blue circles) and 40 kOe (red triangles). d, The multivalued conductivity depending on the cooling field for Pr$_{0.5}$Sr$_{0.5}$MnO$_3$; In this case the measurement fields are zero (blue circles) and 40 kOe (red triangles).}
	\label{fig:Fig2}
\end{figure}

\begin{figure}
	\centering
		\includegraphics{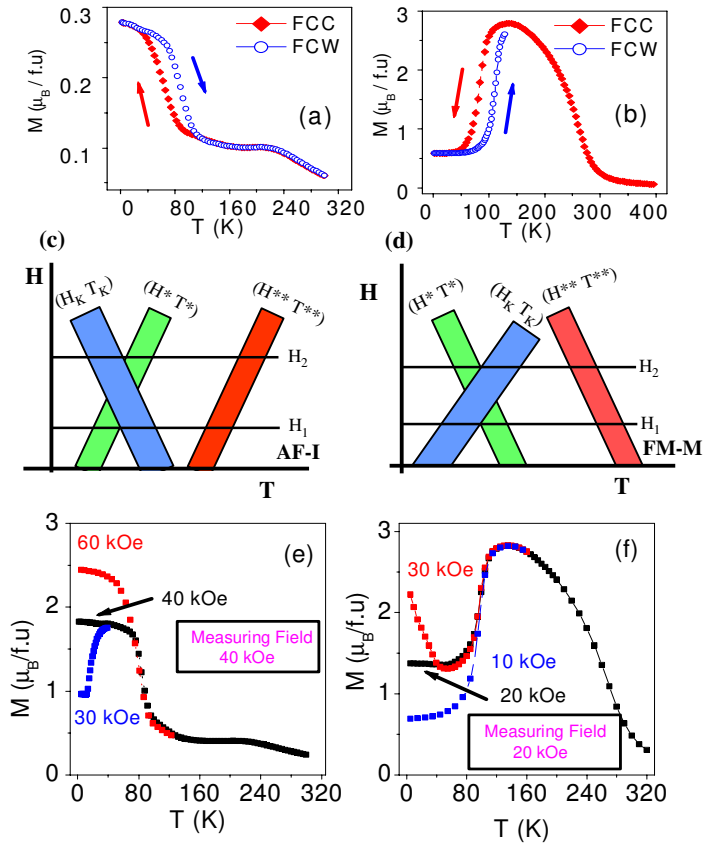}
	\caption{The magnetization of Pr$_{0.5}$Ca$_{0.5}$Mn$_{0.975}$Al$_{0.025}$O$_3$ at 10 kOe is measured during FCC and FCW, and shows a broad hysteretic first-order transition with steep rise at lower temperatures indicating ferromagnetic transformation. The broad hump around 220 K signifies the charge-order transition. b, FCC and FCW magnetization of Pr$_{0.5}$Sr$_{0.5}$MnO$_3$ at 10 kOe shows para- to ferromagnetic transition at high temperature (~250 K) followed by a broad and hysteretic first-order transition from ferro- to antiferromagnetic phase at low temperatures (~120 K). c, The heuristic phase diagram for the case showing transformation from high temperature AF phase to FM at low temperature applicable for the systems like Pr$_{0.5}$Ca$_{0.5}$Mn$_{0.975}$Al$_{0.025}$O$_3$. The various lines are broadened into bands due to disorder. The slope of the bands, as decided from phenomenology, are discussed in the text. d, The heuristic phase diagram for the case showing transformation from high temperature FM phase to AF phase at low temperature applicable for the systems like Pr$_{0.5}$Sr$_{0.5}$MnO$_3$. e, After cooling Pr$_{0.5}$Ca$_{0.5}$Mn$_{0.975}$Al$_{0.025}$O$_3$ in 30, 40 and 60 kOe the field is isothermally changed to 40 kOe at 5 K and the magnetization is measured while warming. The magnetization for higher cooling field of 60 kOe shows one sharp change with the transformation of FM phase to AF phase but the magnetization for the lower cooling field of 30 kOe shows two sharp changes. f, Magnetization of Pr$_{0.5}$Sr$_{0.5}$MnO$_3$ is measured at 20 kOe after following the similar protocol as panel e. Contrary to the previous case the magnetization for the higher cooling field of 30 kOe shows two sharp changes but the magnetization for the lower cooling field of 10 kOe shows one sharp change.}
	\label{fig:Fig3}
\end{figure}

\end{document}